# 10 Gbit·s$^{-1}$ free space data transmission at 9 μm wavelength with unipolar quantum optoelectronics

*Hamza Dely$^+$, Thomas Bonazzi$^+$, Olivier Spitz, Etienne Rodriguez, Djamal Gacemi, Yanko Todorov, Konstantinos Pantzas, Grégoire Beaudoin, Isabelle Sagnes, Lianhe Li, Alexander Giles Davies, Edmund H. Linfield, Frédéric Grillot, Angela Vasanelli, Carlo Sirtori\**

+ These authors contributed equally to this work

H. Dely, T. Bonazzi, Dr. E. Rodriguez, Dr. D. Gacemi, Dr. Y. Todorov, Prof. A. Vasanelli, Prof. C. Sirtori
Laboratoire de Physique de l'Ecole Normale Supérieure, ENS, Université PSL, CNRS, Sorbonne Université, Université de Paris, 24 rue Lhomond, 75005 Paris, France
E-mail: carlo.sirtori@ens.fr

Dr. O. Spitz, Prof. F. Grillot
LTCI, Télécom Paris, Institut Polytechnique de Paris, 19 Place Marguerite Perey, Palaiseau, 91120, France

Dr. K. Pantzas, G. Beaudoin, Dr. I. Sagnes
Centre de Nanosciences et de Nanotechnologies, Université Paris Saclay – CNRS – Université Paris-Sud, 10 Boulevard Thomas Gobert, 91120 Palaiseau, France

Dr. L. Li, Prof. A. G. Davies, Prof. E. H. Linfield
School of Electronics and Electrical Engineering, University of Leeds, Woodhouse Lane, Leeds LS2 9JT, United Kingdom

Free space optics data transmission with bitrate in excess of 10 Gbit·s$^{-1}$ is demonstrated at 9 μm wavelength by using a unipolar quantum optoelectronic system at room temperature, composed of a quantum cascade laser, a modulator and a quantum cascade detector. The large frequency bandwidth of the system is set by the detector and the modulator that are both high frequency devices, while the laser emits in continuous wave. The amplitude modulator relies on the Stark shift of an absorbing optical transition in and out of the laser frequency. This device is designed to avoid charge displacement, and therefore it is characterized by an intrinsically large bandwidth and very low electrical power consumption. This demonstration of high-bitrate data transmission sets unipolar quantum devices as the most performing platform for the development of optoelectronic systems operating at very high frequency in the mid-infrared for several applications such as digital communications and high-resolution spectroscopy.

## 1. Introduction

Unipolar quantum optoelectronics (UQOs) comprises an ensemble of semiconductor devices operating at room temperature in the mid-infrared ($\lambda \sim 4 - 16$ µm) with bandwidths of tens of GHz. The devices exploit the presence of quantum-confined two-dimensional electronic states formed in the conduction band of technological mature semiconductors. They are thus unipolar as electrons are the only charge carriers present. UQOs enables the realisation of optoelectronic systems that combine in phase different devices to produce complex functions and operations.[1] In the mid-infrared this is highly sought after, not only for technological applications, but also to address fundamental physics questions. For example, highly sensitive and ultrafast optoelectronic systems are required for free-space communications,[2–5] light detection and ranging (LIDAR),[6] high resolution spectroscopy,[7] and in observational astronomy.[8,9] On the fundamental side, the coherent assembly of UQOs devices can enable unique experimental arrangements to be devised for quantum measurements, for example to study non-classical state emission from quantum cascade lasers. In this sense UQOs will greatly extend optoelectronic applications and quantum optics into the mid-infrared/THz region.

In this work, we present the realisation of an UQO system for data transmission in the 8 to 14 µm atmospheric window comprising a continuous wave (cw) quantum cascade (QC) laser, an external modulator, and a QC detector (**Figure 1**). In contrast to previous studies based on directly modulating a QC laser current,[2,10–12] data-bits are written onto the laser emission in our system using a high frequency external modulator that operates by shifting the absorption of an optical transition in and out of the laser frequency. This device is designed to avoid charge displacement or electron depletion,[13] and therefore it is characterized by an intrinsically large bandwidth and very low electrical power consumption in comparison with direct current modulation of the laser.[10,14,15] This modulation scheme is a step forward to increase the bitrate for free-space communication with enhanced privacy in the mid-infrared.[16] Using the set-up illustrated in **Figure 1**, we have demonstrated a data rate transmission of 10 Gigabits per second on a single channel, with bit error rate of the order of $10^{-3}$, compatible with common protocols for data transmission. To our knowledge this is the fastest data transmission ever reported in this wavelength range paving the way for commercial communication systems outside the saturated near-infrared telecom bands. In the future, the photonic integration of these devices will further increase their performance and make it possible to extend the realm of quantum technologies deeper in the infrared and THz frequency ranges[17].

**2. System description and high frequency devices**

Our system is sketched in **Figure 1d**. Light from a commercial distributed – feedback (DFB) QC laser (**Figure 1a**) impinges on the modulator (**Figure 1b**) that writes the information, which is subsequently read by the detector (**Figure 1c**). The beam propagation across the modulator is shown in **Figure 1f**: the light coupling into the modulator is through a 60° wedge to increase the coupling length and to facilitate the laser beam alignment. In order to operate at high frequency, the detector and the modulator have been processed into mesa structures that are electrically connected to a 50 Ω coplanar waveguide through an air-bridge for a low - inductance top contact.[14,18] This is shown in the scanning electron microscope (SEM) image in the inset to **Figure 1d**. The device is then fixed on a custom-made holder and wire bonded onto an adapted PCB coplanar waveguide, as illustrated in **Figure 1e**. All devices are realised in III-V semiconductor heterostructures.

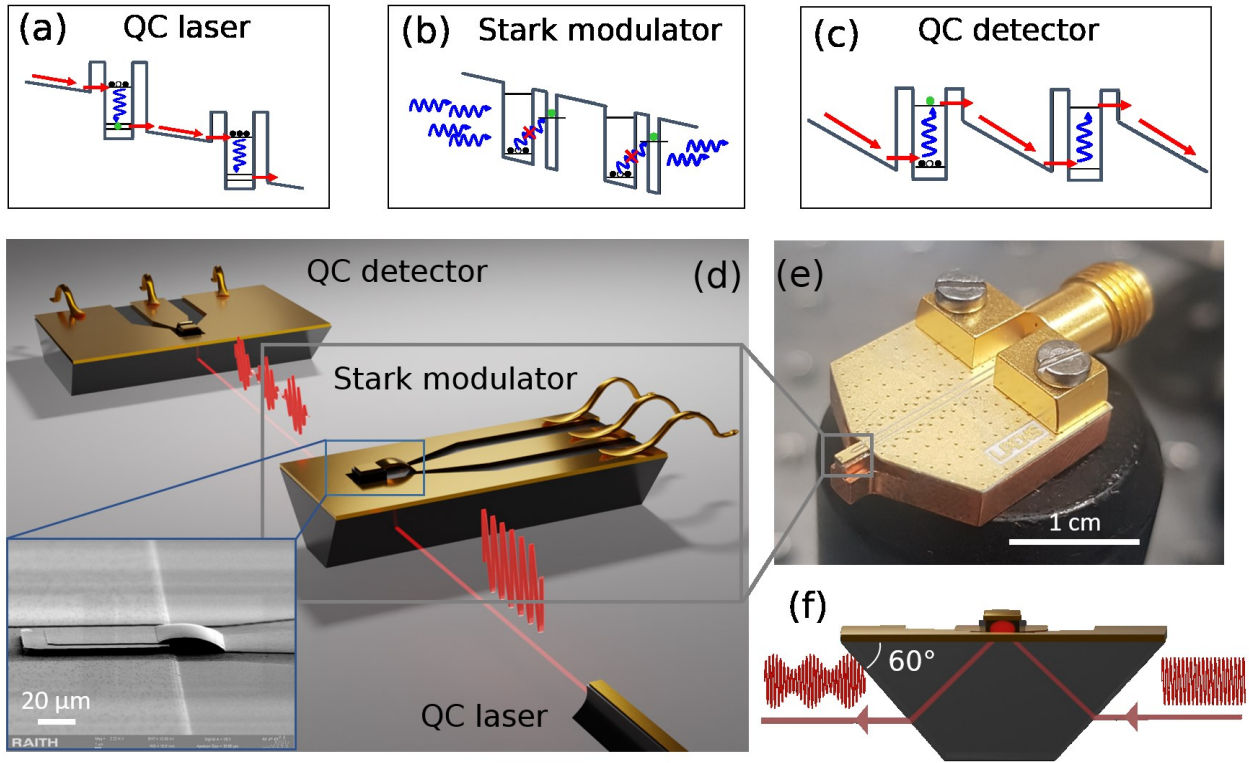

**Figure 1**: (a), (b) and (c) Sketch of the conduction band profile and relevant electronic states of (a) a quantum cascade laser, (b) Stark modulator and (c) quantum cascade detector. Right arrows indicate the electronic path in the structure, while the blue arrows indicate 9µm photons. (d) Sketch of our experiment comprising QC laser, Stark modulator and QC detector, all of them operating at room temperature and at the same wavelength, 9µm (138 meV). The laser is a commercial cw distributed feedback QC laser (Thorlabs QD9000HHL-B). The modulator and the detector have been specially designed, fabricated and mounted to operate at high frequency. The inset shows a SEM image of the modulator connected via an air-bridge to the coplanar waveguide. (e) RF packaging of the Stark modulator. (f) Sketch of the light coupling geometry of the Stark modulator. This geometry complies with polarization selection rules.

The detector is a GaAs/AlGaAs quantum cascade detector, based on a diagonal transition.[19] The geometry for light coupling into the detector is a 45° polished facet, to comply with the polarization selection rules of optical transitions.[20] When photons are absorbed they excite electrons from state 1 to state 2, as sketched in the band diagram of **Figure 1c**. After photoexcitation, electrons relax very rapidly by cascading towards the ground state of the following heterostructure period. This unipolar detector operates in the photovoltaic regime and yet has a very wideband frequency response.[21] This is due to fast energy relaxation of the electrons and the asymmetry of the cascade region that acts as a pseudo electric field driving the electrons in one direction only, giving rise to a photocurrent. The electron relaxation time from one period to the adjacent one is estimated to be shorter than 10 ps and therefore the intrinsic bandwidth is of the order of 100 GHz.[14,18,22] However, the parasitic capacitance induced by the mesa structure limits the frequency range.

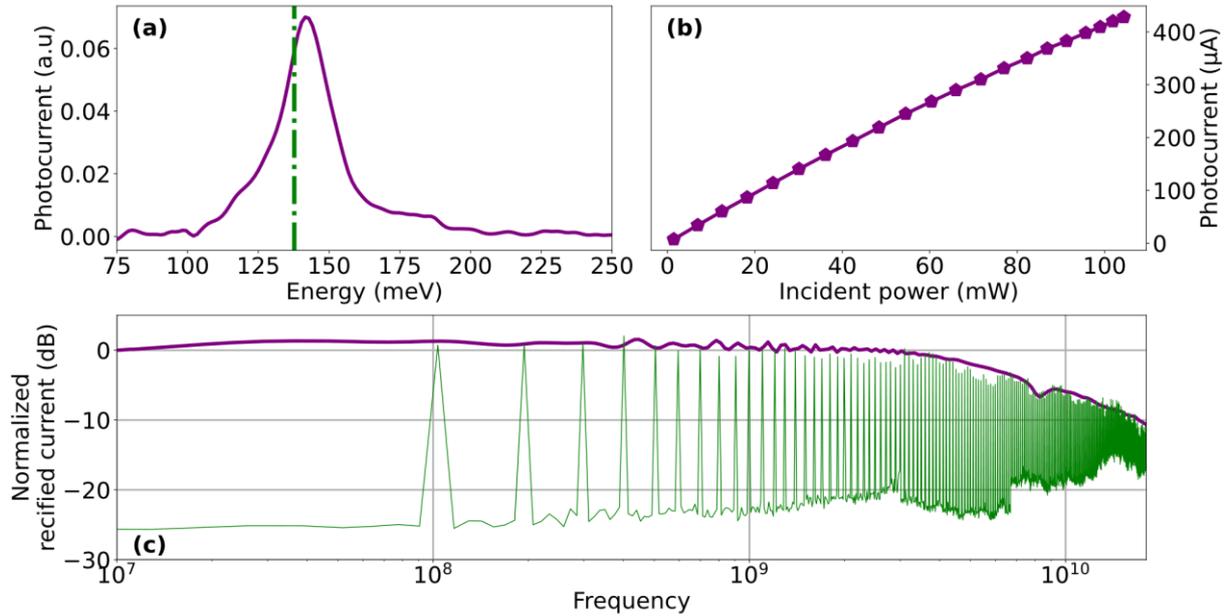

**Figure 2**: Quantum cascade detector characterisation at room temperature. (a) Frequency response of the QC detector, measured by using rectification technique (violet line) and by using a mid-infrared frequency comb (green line) with teeth separated of 100 MHz. (b) Measured photocurrent spectrum at room temperature, centred at an energy very close to the laser emission line (green dashed line). (c) Photocurrent as a function of the incident laser power in cw operation.

The frequency response, shown in **Figure 2a**, is almost flat up to the device cut-off, at 5 GHz for a 50x50 μm² QC detector. Two different methods have been used to measure this response: a rectification technique (violet line) that relies on the non-linear current-voltage (I-V) characteristic,[18] and a direct optical measurement (green line) obtained by shining a mid-infrared frequency comb (Menlo System FC1500-ULN) onto the detector. The beating between the optical teeth appears as beatnotes separated by 100 MHz. **Figure 2b** shows the detector photocurrent spectrum centred at an energy very close to that of the laser emission (green dashed line), while panel (c) presents the photocurrent as a function of the incident cw laser power. The photocurrent is linear with the injected optical power up to 50mW with a responsivity of 4.5 mA·W$^{-1}$. Above 50 mW a mild decrease of the responsivity can be observed, associated with a thermal heating of the detector. This saturation is not a problem in our experiment as the highest optical power on the detector is less than 20 mW.

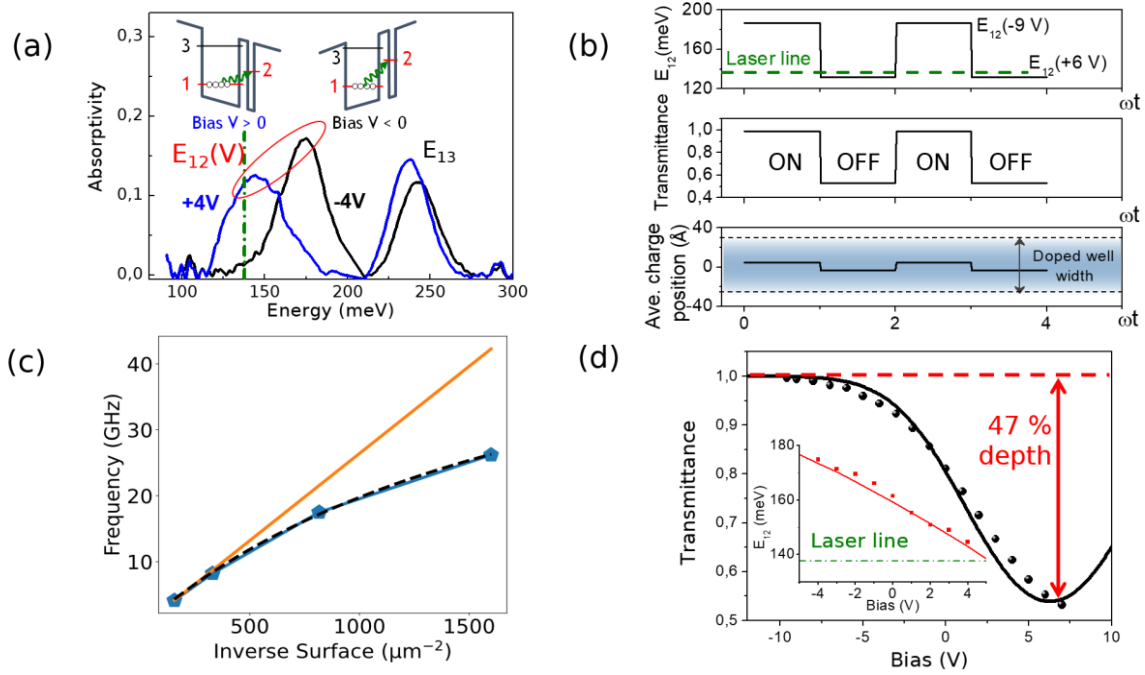

**Figure 3.** (a) Absorption spectra measured at +4V (blue line) and -4V (black line). The green dashed line indicates the laser emission energy. The inset presents a sketch of the conduction band diagram for positive and negative voltage. The bias dependent optical transition involves states 1 and 2. (b) Time operation of the Stark modulator. Top panel: the transition energy $E_{12}$ is modulated by a time dependent applied bias and driven in and out of resonance with respect to the laser emission. Middle panel: time dependent modulated transmittance. Bottom panel: Average displacement of the electronic density. The shaded area indicates the width of the doped quantum well. (c) Cut-off frequency as a function of the device size (symbols), simulated by only including the geometric capacitance of the device (orange continuous line), and simulated considering an extra capacitance of 61 fF due to non-ideal capacitance effects for small-devices (black dashed line). (d) Measured (dots) and simulated (line) transmittance through the modulator as a function of the applied voltage. The inset presents the transition energy $E_{12}$ as a function of the applied voltage, as extracted from voltage dependent absorption measurements (red dots) and through simulation (line).

In order to exploit fully the large frequency bandwidth of our detector for data transmission, we have realized an extremely fast external modulator, based on a linear Stark effect,[23,24] which avoids the implementation of gates for charge depletion, and hence reduces intrinsic parasitic capacitances. The modulator is an asymmetric quantum well[25] made in the GaInAs/AlInAs materials system, *n*-doped at 1.5 x 10$^{18}$ cm$^{-3}$ in the wider well (inset to **Figure 3a** and Methods). This Stark shift originates from the fact that the probability density of electrons in state 1 is essentially localised in the large quantum well, whilst that in state 2 is in the thin well. Under an applied bias $V$, the energy shift of the transition $E_{12}$ equals the drop in potential between the barycentres of the two distributions $L_{bar}$ (which is approximately equal to the distance between the centres of the quantum wells) (see inset to **Figure 3a**):

$$E_{12}(V) = E_{12}(0) + e\Delta V_{qw} \qquad (1)$$

with $\Delta V_{qw} = V \frac{L_{bar}}{L_{struct}}$ and $L_{struct}$ the total thickness of the structure.

A shift of 30 meV can be seen in **Figure 3a** between the low energy absorption peaks of the two spectra measured at +4 V (blue line) and -4 V (black line). Therefore, the absorption at the optical transition $E_{12}$ can be tuned in or out of resonance with the laser emission energy as a function of the applied bias as sketched in **Figure 3b**, illustrating the operation of the device as a function of time. This Stark shift of the $E_{12}$ transition induces a modulation of the laser power absorbed by the device without any charge displacement from the doped well. In fact, the average electronic density is only displaced by a few Angstrom by the change in the applied bias, and lies entirely within the thickness of the quantum well (**Figure 3b** bottom panel). The speed of the modulator is therefore mostly limited by the geometric capacitance of the device, which has been confirmed by performing rectification measurements on devices with different mesa surfaces. The cut-off frequency is shown in **Figure 3c** (symbols) as a function of the inverse of the device surface, together with a simulation (black dashed line) that takes into account an extra capacitance of 61 fF, due to the wire bonding between the coplanar waveguides of the devices and the PCB board, in addition to the geometric capacitance (orange continuous line). The -3dB cut-off frequency is 4 GHz for an 80 x 80 μm² device, and almost 20 GHz for a 25 x 25 μm² device.

The modulation depth has been fully characterised by measuring the absorption under different applied bias, as shown in **Figure 3d**. For a positive bias of +6 V, the transition $E_{12}$ is resonant with the laser photon energy (green dashed line) maximizing the absorption. For negative bias larger than -7 V, the absorption peak is strongly detuned from the laser emission and we obtain the largest transmission. The inset of **Figure 3d** presents the experimental (red symbols) and the calculated (red line) energy of the transition $E_{12}$ as a function of the applied bias. From this plot we can extract a linear Stark shift of 3.9 meV·V$^{-1}$.

The absorption spectrum associated with the 1 → 2 transition can be written as a Gaussian function with voltage dependent peak energy, $E_{12}(V)$, and constant linewidth $\gamma$=17 meV: $\alpha(V) = \alpha_0 \exp\left(-\frac{(E_{12}(V)-E_{laser})^2}{2\gamma^2}\right)$.[23] The transmittance of the structure is $T(V) = \exp(-\alpha(V)L)$ and can be evaluated, by knowing the doping (1.5 x 10$^{18}$ cm$^{-3}$) and the light-matter interaction length $L = 2L_{qw}N_{qw}\frac{\sin^2(\theta)}{\cos(\theta)}$, with $\theta$ the light propagation angle in the device (as sketched in **Figure 1f**), $L_{qw}$ the thickness of the doped quantum well and $N_{qw}$ the number of periods. The resulting voltage dependent transmittance is plotted in **Figure 3d** (black line).

The modulation depth has also been estimated by applying a square signal on the modulator and then measuring the optical power on the QC detector. Using the photocurrent measured when a -9 V bias is applied on the modulator as a reference, we retrieved the transmittance as plotted in **Figure 3d** (black dots), in very good agreement with our numerical estimations. From these data, the modulation depth, $\Delta P = \frac{P_{max}-P_{min}}{P_{max}}$, with $P_{max}(P_{min})$ the highest (lowest) transmitted optical power, is measured to be 47%. Our numerical estimations accurately reproduce these data.

Notably, the modulator shows an excellent linearity in transferring a microwave input to an amplitude signal on the optical carrier, an essential feature for the transmission of analog signals. This has been characterized by analysing the distortion of the optical signal as a function of the amplitude of a sinusoidal input on the modulator $V = V_m \cos(\omega t)$ (see Supporting Information). When the modulator is biased close to the inflexion point of the transmittance as a function of voltage $T(V)$ in **Figure 3d**, the distortion appears as third order sidebands, which allows us to quantify the non-linearity of the modulator. The sideband-to-carrier intensity ratio is over 20dB at 9 V and therefore we can conclude that the modulator operates in its linear regime and does not lead to a significant distortion for a wide range of injected RF power. Furthermore, the QC detector does not introduce distortion as it is operated in its linear regime.

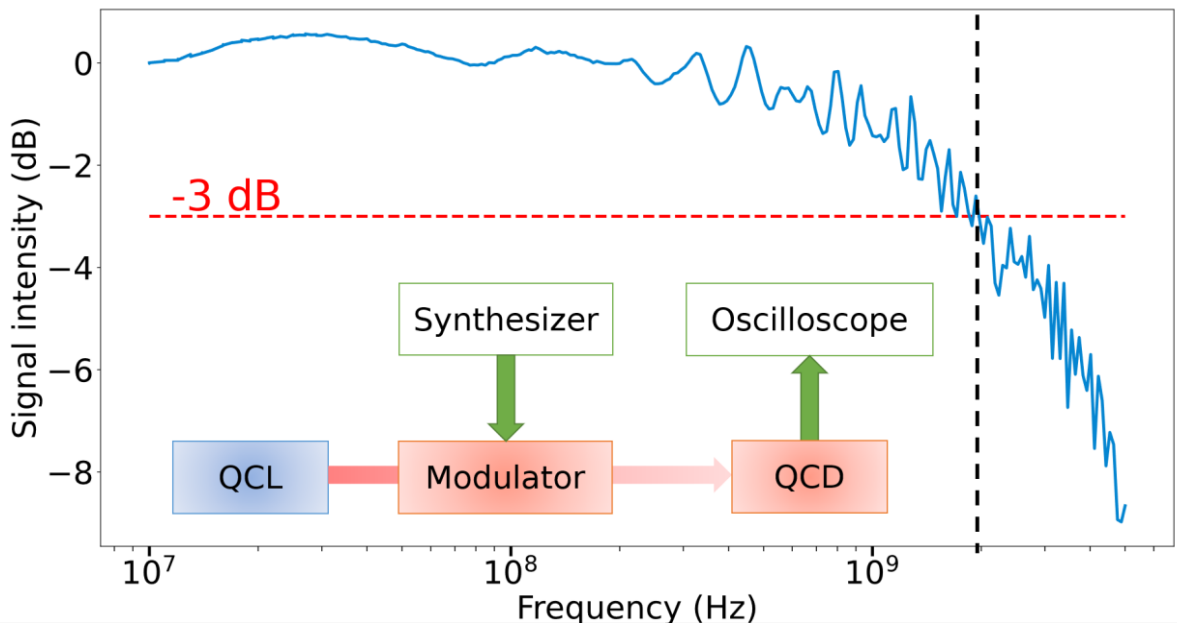

**Figure 4**. Optical frequency response of the full system for free space data transmission. Inset: Sketch of the system. The distance between the QC laser and the QC detector is approximately 2m.

The optical frequency response of the full system for free-space data transmission, which has a cutoff at 2 GHz, is shown in **Figure 4**. The modulator, driven with a power sine wave, writes a signal onto the infrared beam emitted by the QC laser, which is then collected on the detector, and finally is analysed using a 16 GHz-cutoff oscilloscope (Teledyne Lecroy SDA

Zi-B 16 GHz). In the following we illustrate how using these UQO devices we were able to transmit 10 Gbit·s$^{-1}$ with a bit error rate (BER) below 4 x 10$^{-3}$ in free space configuration.

## 3. Data transmission

For the data transmission experiments, the modulator was connected to a pulse pattern generator that outputs 127 bit-long pseudo-random bit sequences (PRBS 2$^7$-1) using a simple on-off keying (OOK) scheme. The latter consists only of 'zeros' and 'ones', so that we have one bit per symbol. The bitrate ranges from 1 Gbit·s$^{-1}$ to 12 Gbit·s$^{-1}$, limited by the pulse pattern generator. The modulated input signal from a random bit sequence at 7 Gbit·s$^{-1}$ and the output of the QC detector on the oscilloscope are shown in **Figure 5a**. The transmission characteristics are analysed using eye diagrams and BER. **Figure 5c** to **Figure 5f** display eye diagrams taken at 7 Gbit·s$^{-1}$ and 11 Gbit·s$^{-1}$. **Figure 5c** and **Figure 5d** are the PRBS references on the oscilloscope, while **Figure 5e** and **Figure 5f** are the corresponding optical eye diagrams of the modulated signal received on the QC detector. Eyes at 7 Gbit·s$^{-1}$ are well-opened, i.e. the ones and zeros are well-resolved, which is typical of a high-quality transmission, while at 11 Gbit·s$^{-1}$ a degradation can be seen from the BER (**Figure 5b**).

The BER presented on **Figure 5b** is obtained by acquiring 100 µs temporal traces for different bitrates and analysing them with an algorithm that we developed ourselves (See Methods). Below 9 Gbit·s$^{-1}$ the signal is error-free. Given that a BER ≤ 4 x 10$^{-3}$ can be corrected using error correction codes without excessive overhead,[2,26] we achieved an error-free transmission up to 10 Gbit·s$^{-1}$ which is far beyond the state-of-the-art at this wavelength in free-space using either external[27] or direct[2] modulation, without any post processing nor equalization.

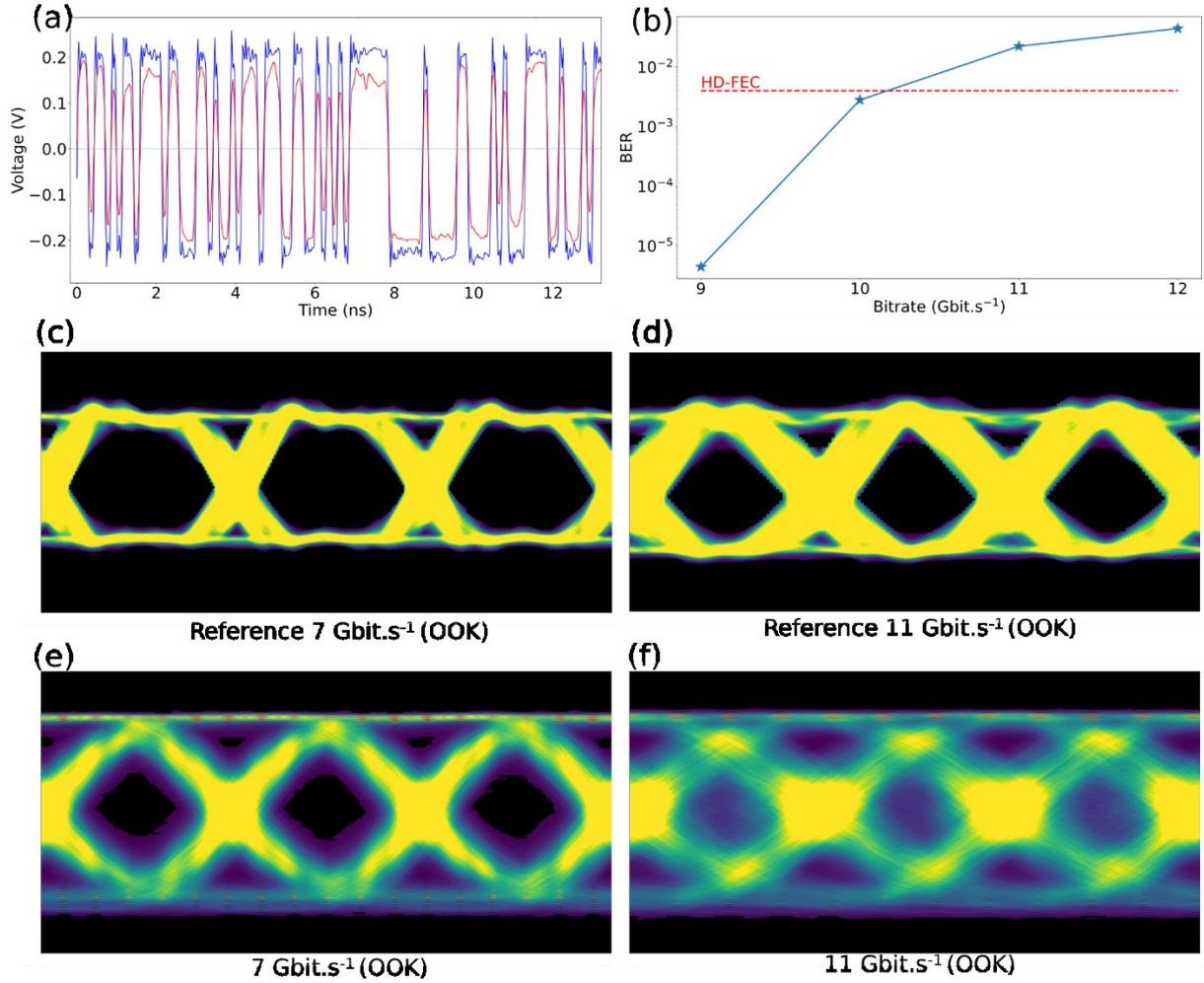

**Figure 5.** (a) Modulated input signal (blue line) from a random bit sequence at 7 Gbit·s$^{-1}$ and output of the QC detector on the oscilloscope (red line). (b) Bit error rate plotted as a function of the bitrate as extracted from a proprietary algorithm (blue line-symbols). The horizontal red line indicates the minimum affordable error rate by using hard-decision forward error correction (HD-FEC). (c) Eye diagram of the reference signal at 7 Gbit·s$^{-1}$ and (d) at 10 Gbit·s$^{-1}$. (e) Eye diagram measured at 7 Gbit·s$^{-1}$ and (f) at 10 Gbit·s$^{-1}$ without any equalization.

## 4. Conclusion

In conclusion, we have realised a data transmission system with an external amplitude modulator that operates in the mid-infrared ($\lambda$ = 9 µm) and outperforms previous results obtained through external or direct modulation of laser current. This has been enabled by ultrafast UQO devices conceived with a multi scale approach from nanometric quantum design to dedicated RF packaging. We have achieved a data-bit rate of 10 Gbit·s$^{-1}$ using an ON/OFF protocol that can be substantially improved further in the present devices, by multi-channel modulation formats, for example through use of discrete multitone (DMT) modulation, and digital processing techniques[28,29], routinely implemented with specific integrated circuits. Moreover, the foreseeable bandwidth of these devices is in the 50 to 100 GHz range and therefore Terabit·s$^{-1}$ data rates should be within reach of this technology, thus positioning the UQO as a possible solution for 6G communications in unregulated frequency bands. Although our results have focused on data transmission, there is a broader scope of applications that arise from more complex UQO systems, involving different devices and different functions such as heterodyne/homodyne detection for instance.

## 5. Methods

*Sample description.* The modulator device is made of $N_{qw}$ =30 periods of InGaAs wells and AlInAs barriers lattice-matched to InP. The thicknesses of the wells and barriers are <u>6.8</u>/**1.6**/2.4/**20** in nm (AlInAs barriers are in bold). The first underlined InGaAs layer is Si-doped to 1.5 x 10$^{18}$ cm$^{-3}$. The quantum wells and barriers are grown by metal-organic chemical vapour epitaxy on a Fe:InP substrate. Growth was performed in a Veeco D180 reactor.

The QCD consists in 12 periods of GaAs wells and AlGaAs barriers. The layer thicknesses in nm are <u>4.4</u>/**1.4**/1.4/**5.5**/1.7/**5.8**/2.3/**5.2**/3.0/**4.8** with the first underlined layer GaAs doped at 1 x 10$^{18}$ cm$^{-3}$. The AlGaAs barriers have a 35% Al concentration and are indicated in bold. The structure is grown by molecular beam epitaxy on a GaAs substrate.

*Fabrication of the high frequency components.* The modulator and the detector were processed into high-frequency coupled mesas in a similar fashion. For the modulator, square mesa regions of 80/55/35/25 µm of side were defined through optical lithography and chemically etched down to the bottom contact layer using $H_3PO_4$:$H_2O_2$:$H_2O$ (1:1:38). Then the half plane starting just below the mesa was protected with resist, allowing the contact to be etched away (using the same $H_3PO_4$ solution) as well as the epitaxial InP buffer (using pure HCl), down to the InP:Fe insulating wafer. The situation following this step is visible in the SEM picture in the inset to **Figure 1d**. After depositing a sacrificial support of reflowed S1818 resist for the air bridge, the Ti/Au 50 Ω coplanar waveguide was evaporated in one step using negative resist AZ5214 E for patterning. A photoresist stripper (SVC 14) was preferred to the standard acetone lift off to ensure that the bridge was properly freed from resist.

The detector underwent the same processing except for the etching step, where the mesa was physically etched (Cl-based ICP). Furthermore, there was no need for buffer etching since the AlGaAs layer below the contact was already insulating.

*Spectra under bias*. Measurements were taken using a Fourier transform infrared spectrometer (FTIR). The sample was processed into 1-mm large ridges to accommodate for the large beam spot, then lapped in a multipass configuration and connected to the electrodes via wire bonding. The sample was cooled to 78 K to limit leakage current under large bias.

*Simulation of the modulator absorption under bias.* The transmittance of the modulator was simulated from Beer-Lambert's law for a Gaussian intersubband absorption spectrum which linearly shifts under bias (3.9 meV·V$^{-1}$ for this Stark shift) : $T = \exp(-\alpha_0 . \exp\left(-\frac{(E_{laser}-E_{12})^2}{2\Gamma^2}\right) * L)$. The absorption coefficient was calculated as $\alpha_0 = \frac{n_{3D}e^2 h}{2\varepsilon_0 c\eta m^*} \frac{f_{12}}{\sqrt{2\pi}\,\Gamma}$ with $n_{3D} = 1.5 \cdot 10^{18}$ cm$^{-3}$ the electron concentration, $\eta = 3.1$ the refractive index, $\Gamma = 17$ meV the linewidth of the transition (considered constant), and $f_{12} = 0.64$ the oscillator strength of the transition 1-2 calculated at 0 V. Finally, the interaction length was taken as $L = 2 L_{qw} N_{qw} * \frac{\sin^2\theta}{\cos\theta}$ with $\theta = 69.3°$.

*Rectification bandwidth measurements.* A sine radiofrequency signal generated with a synthesizer (Anritsu MG3693B) was injected into the device through a wideband bias-tee (SHF-BT45). The non-linear I-V characteristic of the devices generates a rectified DC current $I_{DC} \propto |H(\omega)|^2$ proportional to the squared voltage transfer function $H(\omega)$ of the device. This current was collected from the bias-tee DC connector on a sourcemeter (Keithley 2450).

*Full setup optical bandwidth measurements.* The sine wave synthesizer and a sourcemeter were connected to the modulator through a bias-tee. The same arrangement was performed for the detector with a high-speed oscilloscope (Teledyne LeCroy SDA16 Zi-B) in place of the synthesizer. A temporal trace of the AC signal generated by the detector was acquired for 100 µs at 40 GS·s$^{-1}$ on the oscilloscope and sent to the computer. Finally, a Fourier transform was performed to extract the relevant frequency component.

*Data transfer measurements.* The same arrangement as the one employed for the optical bandwidth was used, replacing the synthesizer with a pseudo-random bit sequence pulse pattern generator (Anritsu MP1763B). We used a 127-bit sequence which is well suited for our system, which presents a lower cutoff at 100 MHz. The quality of the transmission was first analyzed using the oscilloscope data analysis software to obtain the bit error rate, and then a 100 µs temporal trace was acquired and analyzed using a home-made algorithm to compute another estimation of the bit error rate.

*Bit error rate algorithm.* Measurements were performed with a home-made Matlab programme. Given the timetraces from the oscilloscope, this programme resamples the data at a constant sample-per-bit ratio and performs a time-correlation calculation on the input and received signals to calculate the delay between them and realign them accordingly. The sequences of bits associated with each signal were determined and finally compared to each other to get the bit error rate after selecting the most appropriate amplitude threshold.

**Supporting Information**

Supporting Information is available from the Wiley Online Library or from the author.

**Acknowledgements**

We acknowledge financial support from the ENS-Thales Chair, ANR project LIGNEDEMIR (ANR-18-CE09-0035), FET-Open 2018-2020 Horizon 2020 project cFLOW, CNRS Renatech network, and the EPSRC programme grant 'HyperTerahertz' (EP/P021859/1). O.S and F.G wish to thank Dr. Elie Awwad and Andreas Herdt for fruitful discussions.

**References**

[1] Palaferri, D., Todorov, Y., Bigioli, A., Mottaghizadeh, A., Gacemi, D., Calabrese, A., Vasanelli, A., Li, L., Davies, A.G., Linfield, E.H., Kapsalidis, F., Beck, M., Faist, J., et Sirtori, C. (2018) Room-temperature nine-µm-wavelength photodetectors and GHz-frequency heterodyne receivers. *Nature*, **556** (7699), 85-88.

[2] Pang, X., Ozolins, O., Zhang, L., Schatz, R., Udalcovs, A., Yu, X., Jacobsen, G., Popov, S., Chen, J., et Lourdudoss, S. (2021) Free-Space Communications Enabled by Quantum Cascade Lasers. *Phys. Status Solidi A*, **218** (3), 2000407.

[3] Martini, R., Bethea, C., Capasso, F., Gmachl, C., Paiella, R., Whittaker, E.A., Hwang, H.Y., Sivco, D.L., Baillargeon, J.N., et Cho, A.Y. (2002) Free-space optical transmission of multimedia satellite data streams using mid-infrared quantum cascade lasers. *Electron. Lett.*, **38** (4), 181.

[4] Martini, R., et Whittaker, E.A. (2005) Quantum cascade laser-based free space optical communications. *J Optic Comm Rep*, **2** (4), 279-292.

[5] Delga, A., et Leviandier, L. (2019) Free-space optical communications with quantum cascade lasers. *Proceedings Volume 10926, Quantum Sensing and Nano Electronics and Photonics XVI*, 1092617.

[6] Diaz, A., Thomas, B., Castillo, P., Gross, B., et Moshary, F. (2016) Active standoff detection of CH4 and N2O leaks using hard-target backscattered light using an open-path quantum cascade laser sensor. *Appl. Phys. B*, **122** (5), 121.

[7]     Villares, G., Hugi, A., Blaser, S., et Faist, J. (2014) Dual-comb spectroscopy based on quantum-cascade-laser frequency combs. *Nat Commun*, **5** (1), 5192.

[8]     Krötz, P., Stupar, D., Krieg, J., Sonnabend, G., Sornig, M., Giorgetta, F., Baumann, E., Giovannini, M., Hoyler, N., Hofstetter, D., et Schieder, R. (2008) Applications for quantum cascade lasers and detectors in mid-infrared high-resolution heterodyne astronomy. *Appl. Phys. B*, **90** (2), 187-190.

[9]     Hale, D.D.S., Bester, M., Danchi, W.C., Fitelson, W., Hoss, S., Lipman, E.A., Monnier, J.D., Tuthill, P.G., et Townes, C.H. (2000) THE BERKELEY INFRARED SPATIAL INTERFEROMETER : A HETERODYNE STELLAR INTERFEROMETER FOR THE MID-INFRARED. *The Astrophysical Journal*, **537**, 998-1012.

[10]    Paiella, R., Martini, R., Capasso, F., Gmachl, C., Hwang, H.Y., Sivco, D.L., Baillargeon, J.N., Cho, A.Y., Whittaker, E.A., et Liu, H.C. (2001) High-frequency modulation without the relaxation oscillation resonance in quantum cascade lasers. *Appl. Phys. Lett.*, **79** (16), 2526-2528.

[11]    Capasso, F., Paiella, R., Martini, R., Colombelli, R., Gmachl, C., Myers, T.L., Taubman, M.S., Williams, R.M., Bethea, C.G., Unterrainer, K., Hwang, H.Y., Sivco, D.L., Cho, A.Y., Sergent, A.M., Liu, H.C., et Whittaker, E.A. (2002) Quantum cascade lasers: ultrahigh-speed operation, optical wireless communication, narrow linewidth, and far-infrared emission. *IEEE J. Quantum Electron.*, **38** (6), 511-532.

[12]    Martini, R., Gmachl, C., Falciglia, J., Curti, F.G., Bethea, C.G., Capasso, F., Whittaker, E.A., Paiella, R., Tredicucci, A., Hutchinson, A.L., Sivco, D.L., et Cho, A.Y. (2001) High-speed modulation and free-space optical audio/video transmission using quantum cascade lasers. *Electron. Lett.*, **37** (3), 191.

[13]    Pirotta, S., Tran, N.-L., Jollivet, A., Biasiol, G., Crozat, P., Manceau, J.-M., Bousseksou, A., et Colombelli, R. (2021) Fast amplitude modulation up to 1.5 GHz of mid-IR free-space beams at room-temperature. *Nat Commun*, **12** (1), 799.

[14]   Rodriguez, E., Mottaghizadeh, A., Gacemi, D., Palaferri, D., Asghari, Z., Jeannin, M., Vasanelli, A., Bigioli, A., Todorov, Y., Beck, M., Faist, J., Wang, Q.J., et Sirtori, C. (2018) Room-Temperature, Wide-Band, Quantum Well Infrared Photodetector for Microwave Optical Links at 4.9 µm Wavelength. *ACS Photonics*, **5** (9), 3689-3694.

[15]   Hinkov, B., Hugi, A., Beck, M., et Faist, J. (2016) Rf-modulation of mid-infrared distributed feedback quantum cascade lasers. *Opt. Express*, **24** (4), 3294.

[16]   Spitz, O., Herdt, A., Wu, J., Maisons, G., Carras, M., Wong, C.-W., Elsäßer, W., et Grillot, F. (2021) Private communication with quantum cascade laser photonic chaos. *Nat Commun*, **12** (1), 3327.

[17]   Spott, A., Stanton, E.J., Volet, N., Peters, J.D., Meyer, J.R., et Bowers, J.E. (2017) Heterogeneous Integration for Mid-infrared Silicon Photonics. *IEEE J. Select. Topics Quantum Electron.*, **23** (6), 1-10.

[18]   Liu, H.C., Li, J., Buchanan, M., et Wasilewski, Z.R. (1996) High-frequency quantum-well infrared photodetectors measured by microwave-rectification technique. *IEEE Journal of Quantum Electronics*, **32** (6), 1024-1028.

[19]   Reininger, P., Schwarz, B., Detz, H., MacFarland, D., Zederbauer, T., Andrews, A.M., Schrenk, W., Baumgartner, O., Kosina, H., et Strasser, G. (2014) Diagonal-transition quantum cascade detector. *Appl. Phys. Lett.*, **105** (9), 091108.

[20]   Helm, M. The Basic Physics of Intersubband Transitions, in *Intersubband Transitions in Quantum Wells: Physics and Device Applications*, vol. 62.

[21]   Dougakiuchi, T., Ito, A., Hitaka, M., Fujita, K., et Yamanishi, M. (2021) Ultimate response time in mid-infrared high-speed low-noise quantum cascade detectors. *Appl. Phys. Lett.*, **118** (4), 041101.

[22]   Hakl, M., Lin, Q., Lepillet, S., Billet, M., Lampin, J.-F., Pirotta, S., Colombelli, R., Wan, W., Cao, J.C., Li, H., Peytavit, E., et Barbieri, S. (2021) Ultrafast Quantum-Well

Photodetectors Operating at 10 μm with a Flat Frequency Response up to 70 GHz at Room Temperature. *ACS Photonics*, **8** (2), 464-471.

[23]     Teissier, J., Laurent, S., Manquest, C., Sirtori, C., Bousseksou, A., Coudevylle, J.R., Colombelli, R., Beaudoin, G., et Sagnes, I. (2012) Electrical modulation of the complex refractive index in mid-infrared quantum cascade lasers. *Opt. Express*, **20** (2), 1172.

[24]     Teissier, J., Laurent, S., Sirtori, C., Debrégeas-Sillard, H., Lelarge, F., Brillouet, F., et Colombelli, R. (2009) Integrated quantum cascade laser-modulator using vertically coupled cavities. *Appl. Phys. Lett.*, **94** (21), 211105.

[25]     Sirtori, C., Capasso, F., Sivco, D.L., Hutchinson, A.L., et Cho, A.Y. (1992) Resonant Stark tuning of second-order susceptibility in coupled quantum wells. *Appl. Phys. Lett.*, **60** (2), 151-153.

[26]     Ozolins, O., Lu Zhang, Udalcovs, A., Louchet, H., Dippon, T., Gruen, M., Xiaodan Pang, Schatz, R., Westergren, U., Shilin Xiao, Popov, S., et Jiajia Chen (2019) 100 Gbaud PAM4 link without EDFA and post-equalization for optical interconnects. *45th European Conference on Optical Communication (ECOC 2019)*, 257 (4 pp.)-257 (4 pp.).

[27]     Li, T., Nedeljkovic, M., Hattasan, N., Cao, W., Qu, Z., Littlejohns, C.G., Penades, J.S., Mastronardi, L., Mittal, V., Benedikovic, D., Thomson, D.J., Gardes, F.Y., Wu, H., Zhou, Z., et Mashanovich, G.Z. (2019) Ge-on-Si modulators operating at mid-infrared wavelengths up to 8 μm. *Photonics Research*, **7** (8), 828.

[28]     Pang, X., Ozolins, O., Schatz, R., Storck, J., Udalcovs, A., Navarro, J.R., Kakkar, A., Maisons, G., Carras, M., Jacobsen, G., Popov, S., et Lourdudoss, S. (2017) Gigabit free-space multi-level signal transmission with a mid-infrared quantum cascade laser operating at room temperature. *Opt. Lett.*, **42** (18), 3646-3649.

[29]     Chang, F., Onohara, K., et Mizuochi, T. (2010) Forward error correction for 100 G transport networks. *IEEE Commun. Mag.*, **48** (3), S48-S55.

Supporting Information

**10 Gbit·s$^{-1}$ free space data transmission at 9 μm wavelength with unipolar quantum optoelectronics**

*Hamza Dely[+], Thomas Bonazzi[+], Olivier Spitz, Etienne Rodriguez, Djamal Gacemi, Yanko Todorov, Konstantinos Pantzas, Grégoire Beaudoin, Isabelle Sagnes, Lianhe Li, Alexander Giles Davies, Edmund H. Linfield, Frédéric Grillot, Angela Vasanelli, Carlo Sirtori\**

+ These authors contributed equally to this work

***Simulation of the electronic states of the Stark modulator and of the quantum cascade detector***

The electronic states of the heterostructures are numerically calculated by using a three-band Kane model in the envelope function approximation, following reference.[1]

A) <u>Stark modulator</u>

**Figure S1** presents the conduction band profile and the square moduli of the electronic envelope functions, plotted at the corresponding energies. Three electronic bound states are found, whose energies at zero bias are: 95 meV (state 1), 256 meV (state 2) and 359 meV (state 3). Only states 1 and 2 contribute to the absorption of the laser beam at 9 μm, while transition 1-3 is at ~5 μm. The sample is doped $1.5 \times 10^{18}$ cm$^{-3}$ in the large well, so that the Fermi energy is located below state 2 and we can consider that only state 1 is occupied. Electrons are thus mainly localized in the first well, and tunnelling and escape probability are negligible in the entire range of applied bias.

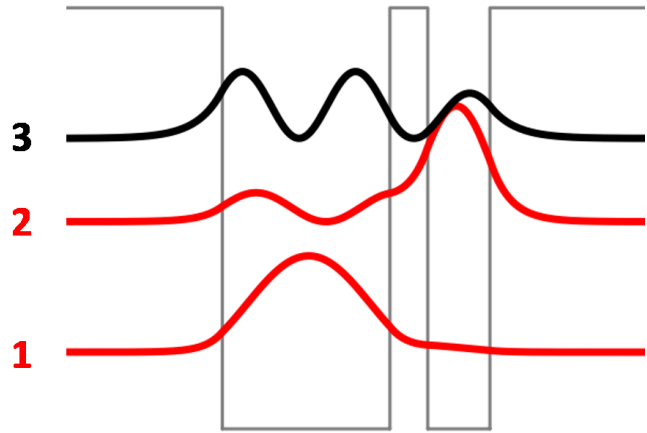

**Figure S1**. Conduction band profile and square moduli of the electronic envelope functions of the Stark modulator.

B) <u>Quantum cascade detector</u>

**Figure S2** presents the conduction band profile of one period of our quantum cascade detector (QCD), together with the square moduli of the electronic envelope functions plotted at the corresponding energies. The design of the active region is based on a diagonal transition:[2] light emitted by the laser photoexcites electrons through absorption between the states 1 and 2, plotted in red. Photoexcited electrons are then extracted through the cascade towards the ground state of the following period by longitudinal optical phonon scattering. This type of design typically yields a lower absorption than its vertical counterpart. The decrease in the oscillator strength is compensated by an improved extraction efficiency through the cascade, thanks to the delocalization of state 2 in the extraction region.

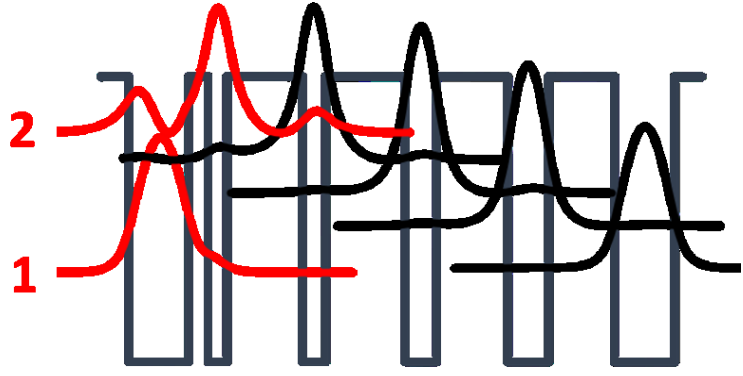

**Figure S2**: Conduction band diagram of the active region of the GaAs/AlGaAs diagonal QC detector, together with the square moduli of the relevant envelope functions plotted at the corresponding energies.

*Linearity of the Stark modulator*

The operation of the Stark modulator was characterized by studying the detected power as a function of the electrical power on the modulator (**Figure S3**). The modulator was operated close to the inflexion point of the transmittance function (**Figure 3d** of the main text), and biased with a DC voltage of 2.05 V and with a sinusoidal voltage $V = V_m \cos(\omega t)$ at a fixed frequency of 12 MHz. The measured photocurrent, from which we extracted the detected optical power, is proportional to the absorbance $\alpha(V)$. Through Taylor expansion of $\alpha(V)$ around the inflexion point, we found that the detected optical power up to the third order in $V$ can be written as:

$$P(V_m) = P_0 \left(1 - m(V_m) \cos(\omega t) + l(V_m) \cos(3\omega t)\right)$$

In this expansion, $m$ is the sideband-to-carrier intensity ratio. The third order allows quantifying the non-linearity of the modulator through the calculation of the input third-order intercept point (IIP3), which was found to be equal to 18 dBm, corresponding to an applied bias of 44 V on our 80 x 80 µm² modulator. Even at large biases covering a large part of the absorption range, this third harmonic lies 20 dB below the fundamental. The input 1 dB compression point (P1dB), where the output signal is 1dB below its linear regime value, is 5.75 dBm (9.5 V). From these observations we can conclude that operating the modulator up to 9 V keeps the device in its linear regime and does not lead to a significant distortion in the modulated signal.

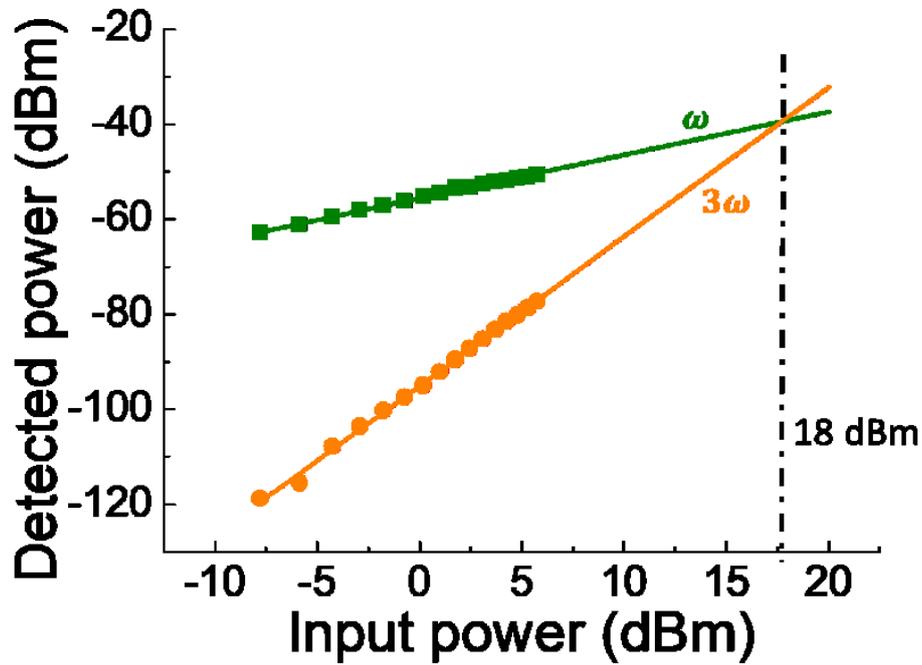

**Figure S3**. Detected power as a function of the electrical power at the modulation frequency $\omega$ (green) and at $3\omega$ (orange).

*Electrical power dissipated by the modulator*

**Figure S4** presents the power dissipated by the modulator as a function of the applied bias, as extracted for each device from the Current-Voltage characteristics measured with Keithley 2450.

For our data transmission experiment, we used a 7 Vpp square signal centered around 1.1 V DC. At this voltage the modulator dissipated approximately 1 pJ/bit. This energy per bit is comparable to state-of-the-art modulation even in the telecom wavelengths.[3]

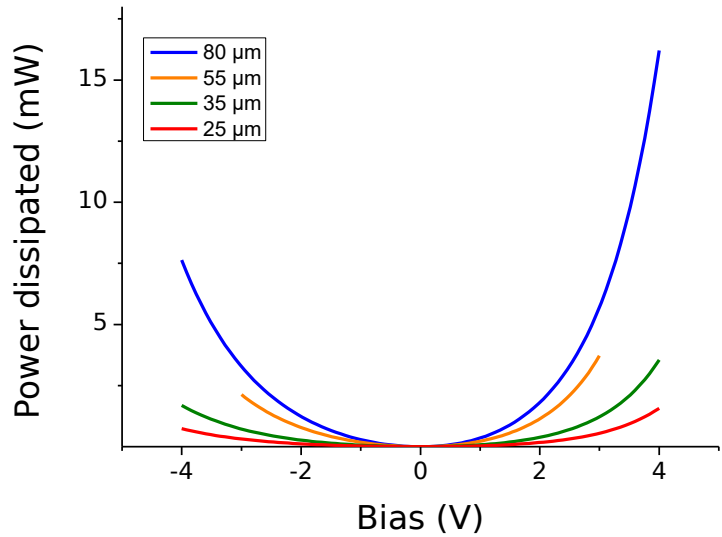

**Figure S4.** Electrical power dissipated by the modulator as a function of the applied bias for four devices with different size of the square mesa.

*Rectification measurements on the Stark modulator*

Rectification measurements were realized on the Stark modulators using a 40 GHz RF probe. This measurement allows extracting the intrinsic bandwidth of the devices without the parasitic capacitance of the PCB mount. **Figure S5** shows the normalized rectified current for four different sizes of the mesa. The four devices exhibit a flat response up to their cutoff and behave like first-order devices (-20 dB/decade) above this point. Note that there is no resonance due to relaxation oscillations near the frequency cutoff, as expected for a unipolar device.

From these measurements we extracted the cutoff frequencies of the four devices, that are plotted in the inset to **Figure S5** (symbols) as a function of the inverse of the device surface. The blue line shows the calculated cutoff frequency only considering the size of the devices, while the orange line includes a parasitic capacitance of 61 fF.

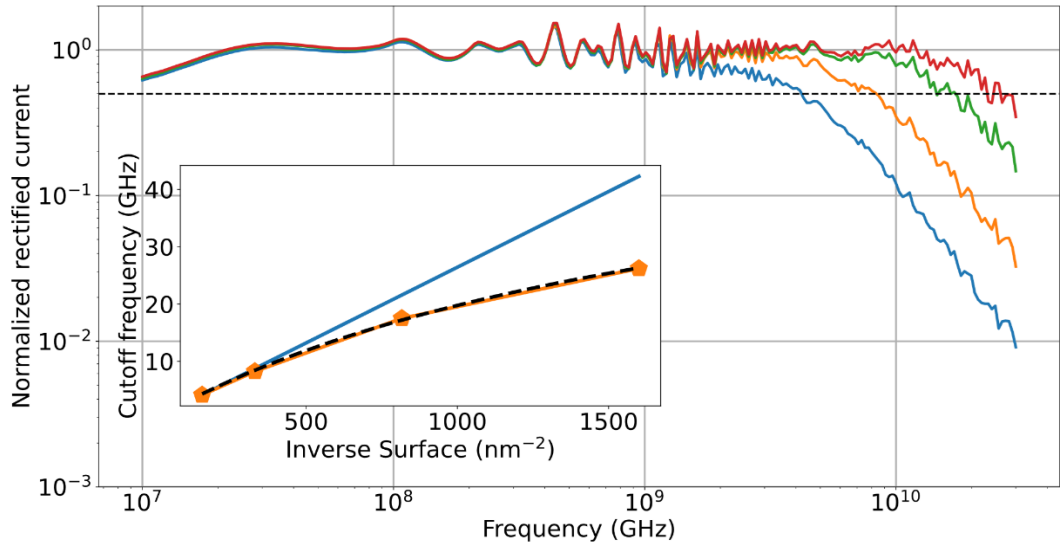

**Figure S5**. Rectification measurements of Stark modulators with different sizes: 80 x 80 µm² (blue curve), 55 x 55 µm² (orange), 35 x 35 µm² (green) and 25 x 25 µm² (red). The inset presents the cut-off frequency as a function of the device size (symbols), simulated by only including the geometric capacitance of the device (orange continuous line) and considering an extra capacitance of 61 fF, due to non-ideal capacitance effects for small-devices (black dashed line).

*References*

1. Sirtori, C., Capasso, F., Faist, J. & Scandolo, S. Nonparabolicity and a sum rule associated with bound-to-bound and bound-to-continuum intersubband transitions in quantum wells. *Phys. Rev. B* **50**, 8663–8674 (1994).
2. Reininger, P. *et al.* Diagonal-transition quantum cascade detector. *Appl. Phys. Lett.* **105**, 091108 (2014).
3. Yamaoka, S. *et al.* Directly modulated membrane lasers with 108 GHz bandwidth on a high-thermal-conductivity silicon carbide substrate. *Nat. Photonics* **15**, 28–35 (2021).